\documentstyle[sprocl]{article}

\input{psfig}
\def\be{\begin{equation}}
\def\ee{\end{equation}}
\def\bea{\begin{eqnarray}}
\def\eea{\end{eqnarray}}
\begin{document}

\title{Classical Approach to Electroweak Dynamics}

\author{ Guy D. Moore }

\address{ Princeton University \\ Joseph Henry Laboratories, 
	PO Box 708 \\ Princeton, NJ 08544, USA }

\maketitle\abstracts{
I discuss the applicability of classical techniques to the study of
the dynamics of infrared, bosonic fields at the electroweak phase
transition.  I present the lattice as a natural means of cutting off
hard, nonclassical modes, and discuss the problem of integrating out
the excluded modes perturbatively.  I then apply the classical lattice
technique to study the dynamics of the phase interface (bubble
wall) and the generation
of Chern-Simons number in each phase and on the bubble wall.  }

\section{Applicability of the classical approximation}

Understanding electroweak baryogenesis requires not only understanding
the thermodynamics of the electroweak phase transition but also the
detailed dynamics of the phase transition, in particular the 
behavior of the out of equilibrium plasma in and around the phase
interface (bubble wall) during the transition.  In particular we are
interested in understanding the dynamics of the Higgs condensate and
of the infrared gauge fields responsible for changing Chern-Simons
number, which through the anomaly is proportional to the creation
of baryon number.  The evolution of these infrared fields cannot be
understood perturbatively because of the infrared problem of thermal
Yang-Mills theory, which has already been discussed in earlier talks.
However there are nonperturbative (numerical) tools available if the
behavior of these fields is well approximated by the classical theory,
as originally suggested by Grigoriev and Rubakov.\cite{GrigRub}
But before using this approximation to study the infrared dynamics,
we should examine in at least some detail whether the approximation
is actually justified.

An absolutely necessary, though not sufficient, condition is that the
thermodynamics are classical.  After all, the thermodynamics are just
the equal time special case of the dynamics.  So I will explore that
question first.

The thermodynamics of the full Standard Model at finite temperature
are governed by the path integral
\bea
Z & = & \int {\cal D}A_{\mu} {\cal D} \Phi \ldots e^{-S/\hbar} \, , \\
S & = & \int_0^{\beta \hbar} d\tau \int d^3 x \bigg[ \frac{1}{4 g^2}
	F^a_{\mu \nu} F^a_{\mu \nu} + (D_\mu \Phi)^{\dagger} (D_\mu \Phi)
	+ V(\Phi^{\dagger} \Phi) + \ldots \bigg] \, ,
\eea
where here and throughout, when I remember, I will write factors of
$\hbar$ explicitly (which seems appropriate if one is studying the
relation between classical and quantum theories), and where $\ldots$
means fermions, gluons, hypercharge (which I leave out for convenience
and because $\Theta_W$ is small), and any other species such as
supersymmetric partners.  These extra species can very easily be
included in what I discuss.

Now the time integral has periodic boundary conditions for bosons and
antiperiodic boundary conditions for fermions, so we are considering a
theory with one compactified dimension.  Like any weakly coupled
Kaluza-Klein theory, we can construct the effective infrared theory
by finding the zero modes of the compact subspace and writing a theory
in the noncompact dimensions with these degrees of freedom.  The 
matching procedure between the full and reduced theories has been
performed by Farakos et. al.\cite{FKRS1,KLRS} as Laine discussed
in his talk.\cite{Lainetalk}  The resulting partition
function is, in a particularly convenient notation,
\bea 
Z_{DR} & = & \int {\cal D}A_{\mu} {\cal D} \Phi e^{- \beta H} \, , \\
H & = & \int d^3 x \bigg[  \frac{1}{4 g^2}F^a_{ij} F^a_{ij} + \frac{1}{2}
	(D_i A_0)^2 + \frac{m_D^2}{2} A_0^2 + \\
	& & + (D_\mu \Phi)^{\dagger} (D_\mu \Phi)
	+ V(\Phi^{\dagger} \Phi) + \frac{g^2}{4} A_0^2 \Phi^{\dagger}
	\Phi \bigg] \, , \nonumber 
\eea
which is exactly the partition function of 3+1 dimensional,
classical Yang-Mills Higgs theory,\footnote{Well, not quite exactly, since
in the classical theory the bare Debye mass is zero, so the value
is forced to equal the counterterm from divergent 1 and 2 loop
graphs.} as first noted by Ambj{\o}rn and Krasnitz.\cite{AmbKras}

Why did that happen?

Perturbation theory is an expansion in nonlinearity, in $g^2$ or
$\lambda$ (though in what follows I will always write $g^2$ in
parametric arguments and estimates).  But $g^2$ has dimensions of 
inverse length times inverse energy.  Since perturbation theory must
be an expansion in a dimensionless quantity (so we can add together the
terms), $g^2$ must always appear with some dimensional quantity which
balances its dimensions.  In the vacuum theory only one quantity has
dimensions involving energy, $\hbar$.  (One frequently thinks of particle
masses, but the classical field theorist would say that the fields have
natural oscillation frequencies and you only get a mass scale by
multiplying by $\hbar$.)  So in vacuum the perturbation theory is an
expansion in $g^2 \hbar$, generally with a $16 \pi^2$ in the denominator
from phase space factors, and $g^2 \hbar / 16 \pi^2 \ll 1$.
Perturbation theory works brilliantly in vacuum, at least for
the electroweak sector.

But at finite temperature there is a natural energy scale in the
problem, the temperature $T$; if there is a length scale $l$ or
frequency $\omega$ either in the question being asked or naturally in
the theory, then $g^2$ can appear in the perturbative expansion in
the combinations
$g^2 l T$ or $g^2 T/\omega$, which need not be $\ll 1$.  So perturbation
theory can be very badly behaved.  However, these combinations do
not contain $\hbar$, so the reason for the breakdown of perturbation
theory is classical.  The expansion in $g^2 \hbar$ works brilliantly,
in vacuum and at finite temperature (the dimensional reduction 
approximation is an expansion in this combination), so while perturbation
theory may go to hell, it will go to hell in a classical way, and 
we can hope to use classical physics to extract the (nonperturbative)
leading order in $g^2 \hbar$ behavior of the system.

The same argument, more or less, works for the dynamics.
If one writes the real time perturbation theory value for any graph,
drops all subleading terms in the Bose-Einstein population function $f_b$,
and replaces $f_b ( \omega ) = T / ( \hbar \omega )$ then one will
recover the classical value for that diagram.  This is a very good
approximation in the infrared, since generally it is $f_b + 1/2$ 
and not $f_b$ which appears and since
\be
f_b(\omega) + \frac{1}{2} = \frac{T}{\hbar \omega} + \frac{1}{12} 
	\frac{ \hbar \omega }{T} + \ldots \, .
\ee
The series has a radius of convergence of $ \hbar \omega / T = 2 \pi$, 
and for small $\hbar \omega/T$ it is approximated
extremely well by its leading term.

However there is obviously a complication.  The classical
approximation is not valid when $\hbar \omega \sim T$, and such
frequencies will inevitably appear in loop corrections even if the
question being considered only involves lower fequencies.  However these
``hard'' degrees of freedom should behave perturbatively and it should
be possible at least in principle to integrate them out and make
a classical theory of the soft modes only.

For the thermodynamics this can be done without too many complications.
Rotational and gauge invariance severely restrict the form of equal time
operator insertions which can appear, and dimensional reduction works.
However, for unequal time dynamics, while vacuum corrections (which
contain all the ultraviolet divergences) are restricted by Lorentz 
invariance, the effects of hard thermal particles need not be Lorentz
invariant, since the plasma chooses a special rest frame.  The form of
these corrections is not so simple.  The dominant terms (in an
expansion in $g^2 \hbar$, which should be well behaved) are the hard
thermal loops.  Physically what they account for is that the infrared
gauge fields at one spacetime point disturb the hard modes propagating
through that point, and that the hard modes carry information about
that disturbance with them to some lightlike separated point in the
future, where they interact again with the soft modes in a way which
conveys information about the fields at the previous location.
This effect is nonlocal, which could be a complication to its
implementation in a regulated classical simulation.  I will come
back to this point later.

\section{Lattice regulation}

We want to study the real time dynamics of the classical theory
nonperturbatively.  The only reliable tools I know of are numerical.
Classical real time lattice gauge theory is particularly well suited
to our needs because it preserves exact gauge invariance.
The lattice implementation restricts the theory to include only 
a finite number of degrees of freedom; in real space, the lattice,
and in momentum space, the first Brillouin zone.  We are thus obliged,
in order to study the classical theory nonperturbatively, to regulate
it in a way which drops precisely those degrees of freedom which do
not behave classically, namely the large momentum or ``hard'' degrees
of freedom.  The problems of separating the soft and hard modes, 
eliminating the hard ones, and nonperturbatively treating the
resulting soft classical modes are thus dealt with simultaneously in an
extremely convenient way.

Ambj{\o}rn et. al. have developed a numerical implementation of the
classical lattice theory\cite{Ambjornetal} and both Alex Krasnitz
and I have developed thermalization algorithms.\cite{Krasnitz,Moore1}
I will not discuss the details of the implementation here, 
because it has already been presented at this conference,\cite{Krastalk} and
because you either already know, don't want to know, or
would be best advised to learn directly from the literature.

I should also note that the classical evolution probes a system
with the same thermodynamics as the quantum theory in the dimensional
reduction approximation, and because the classical evolution is ergodic,
it can be viewed as a microcannonical monte-carlo of the dimensionally
reduced theory, which has some advantages over conventional cannonical
approaches--in particular, mixed phase configurations are stable and
naturally adjust to find the equilibrium temperature.\cite{MooreTurok}  
But I will not go into detail on this matter here.

Instead I should discuss at a little more length the problem of 
integrating out the hard modes which are present in the continuum
theory, rather than simply dropping them.  Again, the first step
should be to demand that the thermodynamics of the full system are
recovered with the smallest possible error.  This is best done
in a two step process.  First, one goes from the full quantum theory
to the dimensionally reduced one, which is the same as studying the
thermodynamics of the continuum classical system.  This step has
been carried out at one loop by
Farakos et. al.\cite{FKRS1}$^,$\cite{KLRS} and discussed
in Laine's talk.  Next, one should make contact between the continuum
3-D theory and the lattice one.  This problem has also been studied
by Farakos et. al.\cite{FKRS}$^,$\cite{Laine}, 
who show that to make the continuum
theory and the small $a$ (lattice spacing) limit of the lattice theory 
match, one need only make one and two loop renormalizations of the
Higgs and Debye mass squared and of divergent operator insertions
such as the $\Phi^2$ insertion.  They compute these corrections, yielding
a system for which the small $a$ limit rigorously exists and matches
the continuum theory.

This does not address the question of how quickly the theory approaches
this small $a$ limit, which is of some practical importance in numerical
lattice studies.  In particular we do not want there to be any $O(a)$
differences between the lattice and continuum theories.
Super-renormalizability is very useful here as well.
The continuum and lattice theories differ, at tree level, 
at $O(a^2)$ in the infrared,
and so $O(a)$ errors in infrared phenomena
can only arise from diagrams with hard loops, ie
where some propagators carry momenta on order $1/a$, since only such
propagators (and the verticies they go into) differ substantially
between the lattice and continuum theories.
The superficial degree of divergence of any 3-D Feynman diagram is 
smaller by 1 for each loop than the corresponding 4-D graph, which is
why divergences could only happen in 1 and 2 loop corrections to the mass.
It also means that corrections to couplings and wave functions, which
are normally logarithmically divergent, are now convergent by one power
for each loop order.  Order $a$ corrections will occur due to hard
momenta in isolated loops which, if contracted, would appear as coupling
or wave function corrections.  Overlapping hard loops will differ between
the theories at $O(a^2)$ (except for mass squared corrections, as I
mentioned) and need not be considered; and corrections other than wave
functions and super-renormalizable couplings only correct nonrenormalizable
operators which are already $O(a^2)$.  Hence a 1 loop integration over
the modes missing on the lattice can remove all $O(a)$ errors, except in
the mass squared and in divergent operator insertions.  Neither of
these exceptions are very important,
since we tune the mass to find the phase transition, and
since we are generally interested in the difference of operator
insertions between the two phases, and these are convergent.

\begin{figure}[t]
\centerline{\mbox{\psfig{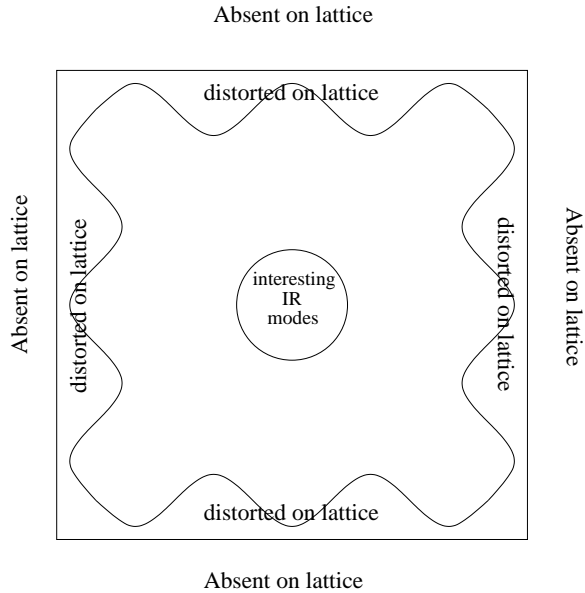}}}
\caption{Illustration of the difference between the lattice and continuum
theories, in terms of the degrees of freedom in momentum space.  The
lattice theory only contains degrees of freedom inside a Brillouin 
zone; the degrees of freedom outside are dropped, and the behavior of
the degrees of freedom near the edges are distorted.  The interesting
degrees of freedom are the infrared modes around the origin.
\label{figOa}}
\end{figure}

A picture here is worth several words.  Figure \ref{figOa} shows how
the lattice only contains some of the degrees of freedom in momentum
space.  Since we are interested in the IR degrees of freedom around
$p=0$, this is only important in that the UV degrees of freedom interact
with them via hard loops.  Because the theory is super-renormalizable,
the coupling is weaker and weaker as we go further from the origin; the
degrees of freedom we drop are weakly coupled.  The finer the lattice,
the larger the box in the figure and the fewer and more weakly coupled
the modes we lose; so a small $a$ limit exists.  But the convergence
to the limit is much faster by performing a one loop calculation of
the influence of the hard modes, on the lattice and in the continuum, and
applying counterterms equal to the difference to make up for the absence
of the hard degrees of freedom.  The matching has been 
performed,\cite{Oapaper} and an algebraic error in the original draft
has been corrected.

What about integrating out the hard modes in a way which will be 
correct dynamically as well?  As I mentioned earlier, these modes
propagate at the speed of light and cause interactions between lightlike
separated points.  Reproducing those effects correctly on the lattice
is tricky.  There are now three proposals which I know 
of \cite{Smilga}$^{\!-\,}$\cite{Muller} but none has been implemented
and used in the existing literature and I think it would be fair to say
that the subject is immature.

However, there are already ``hard'' modes, in the sense of short wavelength
excitations which interact weakly with the infrared modes and carry
almost all of the energy in the system, present on the lattice; namely,
the high frequency lattice excitations.  They propagate in the background
of the soft lattice modes and lead to interactions quite similar to
the hard thermal loops.  For instance, they lead to damping of the same
sort as the hard loops.\cite{Arnoldnew}  However, since their dispersion
relations are not ultrarelativistic and their population cuts off in
a way which is not rotationally invariant, the hard thermal loops they
contribute look different from the correct ones.\cite{Smilga}

Does this difference matter?  Probably yes.  As Peter Arnold argued
nicely in his talk, the presence of hard thermal loops slows down the
dynamics of very infrared magnetic fields in a way which depends on
how many such modes there are and on how exactly they modify the 
self-energy of the infrared modes.\cite{ArnoldYaffe}  
The lattice theory at some arbitrary lattice spacing will typically get
this wrong, leading to the wrong strength of damping on the infrared
magnetic fields, which could change the Sphaleron rate.  
But for exploratory purposes
I will start out relying on the lattice modes to convey the hard thermal
loops, and look on the results not as high precision measurements but
as qualitative, intended as much to develop techniques as to determine
hard numbers.  When the problem of including hard thermal loops more
accurately has been solved, which I think will be in the relatively 
near future, then we can apply these methods to get hard answers (when
we know the right electroweak physics, which is another reason not
to be looking for final answers at this point in the game).

\section{Bubble wall friction}

With that said, I will turn to an example of computing an interesting,
dynamical infrared quantity with the classical field technique.  Namely,
I will compute the friction on a moving bubble wall, below the phase
transition temperature.

At equilibrium, the interface between the two electroweak phases will
on average remain at rest.  (That is what we mean by equilibrium.)  But
when the temperature is lowered, the low temperature ``broken'' phase
has a lower free energy density, and thus a higher pressure.  There is
a pressure difference across the interface which pushes it, allowing the
broken phase to expand at the expense of the symmetric phase.  The
problem of electroweak baryogenesis is to understand in detail what 
happens on and near this moving bubble wall, and how fast the wall
moves is an important attribute.

The velocity is set by a balance of the velocity independent net pressure
on the bubble wall and the velocity dependent frictive force as it
moves through the plasma.  I define a friction coefficient for
the bubble wall as the linear response coefficient relating the
mean velocity $v_w$ to the net pressure $P$,
\begin{equation}
\eta \equiv \lim_{T \rightarrow T_c} \frac{P}{v_w} \, ,
\end{equation}
which should not be confused with the $\eta$ defined in Ignatius' talk.

The first thing to note about $\eta$ is that there is a one loop
perturbative estimate,\cite{Arnoldwall} which is 
equivalent to the friction in the free scatter approximation derived
independently by Dine et. al.\cite{Dineetal} and Liu et. al.,\cite{Liu}
and which for the bosons is strongly infrared dominated.  This tells
us at once that the contribution cannot be found reliably in
perturbation theory, but that it is classical.  The second thing to note
is that we can find this linear response coefficient from an unequal time,
equilibrium correlator via a fluctuation dissipation relation, similar
to and inspired by one developed by 
Khlebnikov.\cite{Khlebwall}

The idea is the following.  Consider the bubble wall in a long box of
cross section area $A$.  Once we choose an unambiguous definition of the
wall position, it is a coordinate, and it has a conjugate momentum.  
Equipartition tells us that it will typically be moving.  But as it
interacts with the plasma, its momentum will repeatedly receive kicks
which will change its direction; so it will diffuse with a diffusion
constant \footnote{One can prove that the position will diffuse, assuming
only ergodicity and translational invariance.}
\be
D \equiv \lim_{t \rightarrow \infty} \frac{ \langle (x(t) - x(0))^2 
	\rangle}{t} \, .  
\ee
The more kicks it
receives, the more slowly it will diffuse, because the individual steps
making up its diffusive motion will be shorter.  The kicks it is receiving
are precisely the kicks which will absorb the net motion generated by
a pressure in the out of equilibrium case, so the friction coefficient
and the diffusion constant are inversely related.  In fact, a simple 
argument gives
\be
\eta = \frac{2 T}{D A} \, .
\ee
Except for the factor of 2 one can guess this equation just on dimensional
grounds; $\eta$ must go as energy density over velocity, that is
$[\eta] = {\rm energy}\times{\rm time}\times{\rm length}^{-4}$.
Now $T$ is the only unit of energy in the classical theory, $A$ has units
of length$^2$, and $D$ has units of length$^2 /$time.  The 2 is the same
2 which appears with $T$ in the noise correlator in Langevin equations
and which appears in the fluctuation dissipation relation between 
Chern-Simons number diffusion and motion under a chemical potential.

The fermionic contribution to the friction was computed, in an approximation
which should be correct at least parametrically, by Tomislav
Prokopec and me,\cite{MooreProkopec} who found $\eta \sim \hbar g^8 T^4$.  
This contribution might be relevant because four of the $g$ are top
quark Yukawa couplings, but as we will see it is parametrically 
smaller than the classical bosonic contribution.  The naive parametric
estimate for the classical contribution is that the only classical
length scale is $1/(g^2 T)$, which is also the only classical time
scale; so $\eta \sim g^6 T^4$.  Peter Arnold might disagree with this,
though.  The two electroweak phases are distinguished not only in that
one has a larger value of $\phi^2$, but also that the ``symmetric'' phase
has large IR magnetic fields, while the broken phase does not.  For the
wall to move, it must create or destroy these large magnetic fields,
and as Arnold, Son, and Yaffe argue,\cite{ArnoldYaffe} 
the time scale for these to change
is $1/(\hbar g^4 T)$ due to HTL damping effects.  That would give
an estimate for $\eta$ of $\eta \sim g^4 T^4/\hbar$.  If this reasoning
is right, then the fermionic contribution to the damping is clearly
quite irrelevant.

It is straightforward to measure the diffusion constant for the wall
surface numerically.  Neil Turok and I did so; our results, applying 
the $O(a)$ corrections which I mentioned earlier, are that, for
$4\lambda/g^2 = 0.159$ (or $ x \simeq 0.04$), on a lattice with
$\beta_{L,imp} = 7.3$, we find $\eta = 0.020 \pm 0.003 g^6 T^4$.
At these parameter values, we found a jump in the order
parameter of $\Delta \phi^2 = 2.11 g^4 T^2$,
which is right on the 2 loop perturbative value, and a surface
tension of $\sigma = 0.057 g^4 T^3$, which is somewhat below the perturbative
value, a trend which gets stronger at larger $x$.  If we use the value
of $\Delta \phi^2$ to compute the ``free scattering'' estimate of the 
friction coefficient, we get $\eta = 0.058 g^6 T^4$, which is larger.
However, the strength of hard thermal loop effects, induced by the hard
lattice modes, is much smaller here than in the physical, quantum theory, and
if the Arnold Son Yaffe argument is right, the true friction, when this is
taken into account, should be larger than the one we measured
by a factor of 3 or 4.  That is, the friction could 
be larger than the free scattering estimate, which will definitely
mean that the bubble wall moves slowly, $v_w \sim 0.1$ at the
nucleation temperature (and still slower after the universe starts to 
reheat towards the equilibrium temperature).

Hence we conclude that bubble walls are slow, although this might 
be different for extremely strongly first order transitions.  But more
work, and in particular a better accounting of the hard thermal loop
effects, is needed before we can quote a solid number.

\section{Dynamics of Baryogenesis}

If we can find a way to measure the (topological) quantity $N_{CS}$ on
the lattice then we can study how it changes in circumstances which are
relevant to baryogenesis.

A local operator (non-topological, non total time derivative)
definition of $N_{CS}$ has been developed by Ambjorn and 
Krasnitz,\cite{AmbKras} and is very convenient for lattice measurements.
There are three interesting problems we might try to address with this:
\begin{itemize}
\item What is the diffusion constant of $N_{CS}$, or the 
	linear response coefficient to a chemical potential for
	$N_{CS}$, in the symmetric phase?
\item Same, but for the broken phase.
\item What about on the bubble wall, say, while it is moving and out of
	equilibrium?
\end{itemize}
A nice fluctuation dissipation 
relation \cite{KhlebShap}$^,$\cite{RubShap}$^,$\cite{Moore1} tells us that 
the diffusion constant,
\be
\Gamma_d \equiv \lim_{t \rightarrow \infty} \frac{ \langle ( N_{CS}(t) - 
	N_{CS} ( 0 ) )^2 \rangle }{ V t } \, ,
\ee
and the linear response coefficient,
\be
\Gamma_\mu \equiv \lim_{\mu \rightarrow 0} \frac{ T \langle \dot{N}_{CS} 
	\rangle_{\mu} }{V \mu} \, ,
\ee
are related, $\Gamma_d = 2 \Gamma_\mu$.  So the first two questions can
either be solved by the diffusion constant technique developed by
Ambjorn et. al.\cite{Ambjornetal} and applied by Ambjorn and 
Krasnitz \cite{AmbKras} and Smit and Tang,\cite{TangSmit} or by the 
chemical potential technique.\cite{Moore1}$^,$\cite{MooreTurok}  The third
question can only be addressed by the chemical potential technique, and 
I will address it next.

\begin{figure}
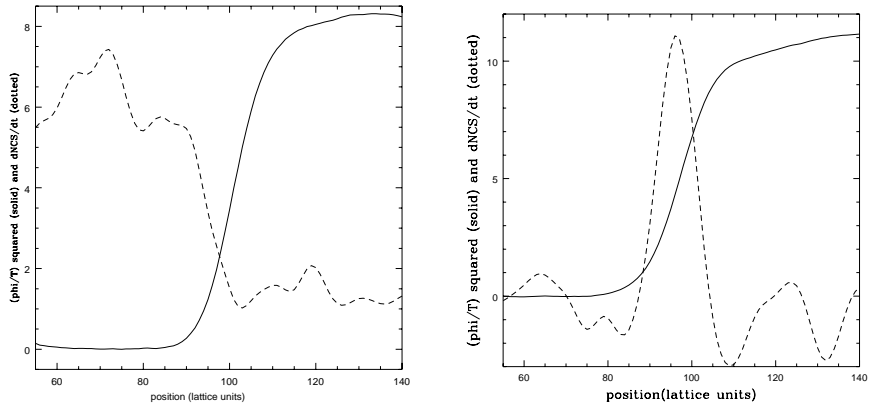

\centerline{
\mbox{\psfig{file=atrest.epsi,width=2.1in}} \hspace{0.2in}
\mbox{\psfig{file=muonwall.epsi,width=2.1in}}}
\caption{At right, generated $N_{CS}$ when a chemical potential is
applied uniformly throughout the box; at left, the same when the chemical
potential is only applied in the interior of the wall.  For the uniform
application, the generation falls off at the leading edge of the wall.
For the application in the wall's interior, some $N_{CS}$ is generated,
but there is a trough on either side of the generation region and
the net production is quite small.  The horizontal scale is quite
close to units of $\hbar/T$, the vertical scale for $\phi^2$ is
in units of $g^2 T^2 / 4$, and the scale for $N_{CS}$ is 
arbitrary.  \label{NCSfig}}
\end{figure}

We want to study how the system's response to a chemical potential
for $N_{CS}$ changes from the symmetric phase, unsuppressed rate to
the broken phase, exponentially suppressed rate at the phase interface,
a question which is relevant for ``local'' mechanisms for baryogenesis.
The first thing we did was to get a mixed phase configuration in a very
long box and to apply a chemical potential uniformly through
space, and measure the generation of $N_{CS}$ as a function of distance
from the bubble wall.  The result is the picture on the left in Figure
\ref{NCSfig}.  We see that the generation of $N_{CS}$ shuts off just
past the base of the bubble wall.  (There is some slight generation in
the broken phase, but this is a lattice artifact associated with this
definition of $N_{CS}$; a spurious response in the broken phase of about
$10 \%$ of the symmetric phase rate is generated by UV lattice artifacts.)
The picture for the out of equilibrium case, where the system
is cooled and the wall propagates, is about the same.

The other thing to try is to apply a chemical potential only to the 
interior of the bubble wall.  To do this, we first measure the 
averaged dependence of $\phi^2$ on distance from the bubble wall, which
appears already on the lefthand picture in Figure \ref{NCSfig}.
Then, at each point in time, we find the bubble wall surface, and determine
the vertical distance of each point in the plasma from the bubble wall,
and apply a chemical potential at that point which is proportional to
the derivative of the wall profile.  The chemical potential is then only
nonzero in the interior of the bubble wall, in a way which approximately
duplicates what would happen in a ``local baryogenesis'' mechanism.

The result of this endeavor is shown in the righthand picture in 
Figure \ref{NCSfig}.  Integrating over volume, we find a net production of
$N_{CS}$ which is about $20 \%$ of the value we would get by applying the
same integrated chemical potential in the symmetric phase.  Remembering
that half of this is the UV lattice artifact, and that the wall has
a breathing mode so it is thinner some places than others (resulting in
some small part of the chemical potential getting applied directly in the
symmetric phase), we can set an upper limit on the efficiency of baryogenesis
inside the bubble wall of $10 \%$ of the symmetric phase efficiency.
This is bad but not fatal for local baryogenesis mechanisms, ie it
appears that transport is important to make baryogenesis efficient.

\section{Conclusions--What needs doing?}

\begin{figure}[t]
\centerline{\psfig{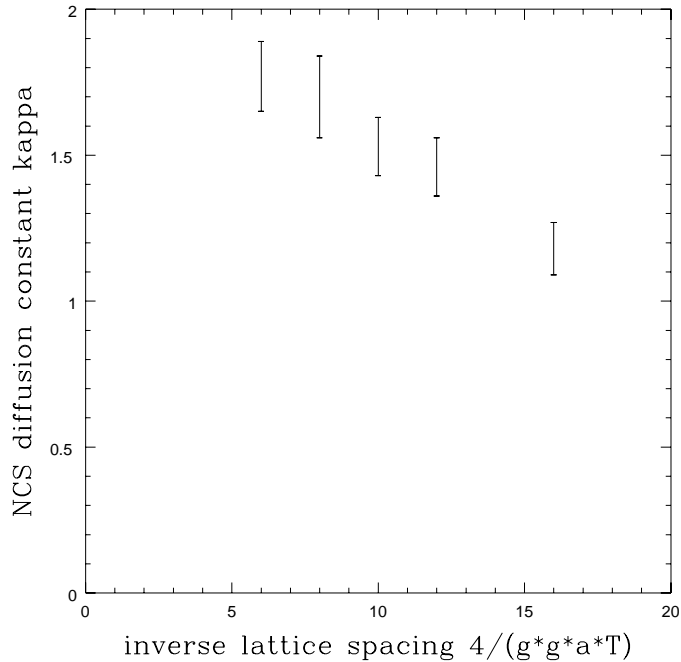}}
\caption{Lattice spacing dependence of the diffusion constant for
$N_{CS}$ in pure Yang-Mills theory, using a topological 
definition for $N_{CS}$ and after applying $O(a)$
corrections in the relation between lattice and physical length 
scales. \label{result}}
\end{figure}

I see three serious drawbacks with the study of $N_{CS}$ violation I
have just discussed.  The first is with the definition of $N_{CS}$,
ie using a local operator which is not a total time derivative and
which has UV noise.  

This problem has now been solved, as Alex Krasnitz
reviewed in his talk.  Neil Turok and I have developed a topological 
means to track $N_{CS}$,\cite{slavepaper} based on a topological
definition of lattice $N_{CS}$ due to Woit.\cite{Woit}  With it, we
have studied the lattice spacing dependence of the diffusion constant
for $N_{CS}$, and find a strong spacing dependence, as shown in
Figure \ref{result}.  (Note that we have already applied the $O(a)$
corrections in the match between lattice and physical length scales to
the data in the figure; before the correction the lattice spacing
dependence is stronger.)  We expect a lattice spacing dependence if Arnold,
Son, and Yaffe are correct, because the total size of HTL effects depends
linearly on $1/a$ (as more and more hard modes are present on the lattice).
The size of the effect we find is not as strong as their prediction, but
this could be because the parametric limit of small $a$ has not yet been
completely attained.

Another problem, which should be solved rather than applying the above
methodology to finer and finer lattices, is the problem of more correctly
including hard thermal loop effects on the lattice.  I know of three
proposals in the existing literature for doing this.

The first proposal is that
of B{\"o}deker, McLerran, and Smilga,\cite{Smilga} which consists of having
a population function defined on a sphere at each lattice site, representing
the population of particles moving in each direction, and to evolve these
population functions according to lattice Boltzmann equations in the
presence of the classical lattice background.  The specifics of the
lattice implementation have not been worked out and it is not clear
to me if they would be practical.  

The second proposal is that of
Huet and Son,\cite{HuetSon} who propose to treat directly the 
nonlocal dynamical system resulting from integrating over the hard modes,
but to make all available parametric approximations to simplify the system.
The result is a set of Langevin equations which are nonlocal in space
but not in time.  The idea has been formulated in the continuum, but
carrying it over onto 
the lattice is straightforward; the only problem is dealing
with a highly nonlocal update rule, which is numerically impractical without
either a very clever trick or some additional approximations.

The third proposal is that of Hu and M{\"u}ller,\cite{Muller} who
propose to include the hard thermal loops in a local way by adding particles
to the lattice, with dynamics which reproduce the continuum Wong's equations
in the small spacing limit.  A numerical implementation appears feasible,
although there are some systematics and orders of limits to be carefully
taken care of.

I think that one or the other of these proposals, or perhaps a new proposal,
for dealing with the hard thermal loops numerically, will be implemented
in the foreseeable future, and I will make a guess that, when one
has been, we will find a $N_{CS}$ diffusion constant, for HTL strengths
corresponding to the Debye mass of the real quantum theory, of
$\Gamma_d = (0.4 -- 0.9) \alpha^4 T^4$, with $0.6$ being my best bet.

Finally, it would be nice to redo the study of $N_{CS}$ generation ``on the
bubble wall,'' but introducing the $CP$ violating bias in the evolution
of $N_{CS}$ through a nonrenormalizable operator self-consistently
included in the lattice theory.  Implementing this would be a real pain,
but there is no problem in principle.

About the computation of the friction on the bubble wall, obviously the
most imporant thing to do here is to apply the hard thermal loops in a
more realistic way.  It would also be nice to have results for a 
variety of values of the scalar self coupling.  And realistically, we
expect that there must be new light bosonic degrees of freedom in the
theory if the phase transition is to be strong enough for baryogenesis
to work.  Friction from these extra degrees of freedom depends on the
strength of their coupling to the Higgs field, so in particular we might
expect a large contribution to the friction in the case that there are
light stops.

And speaking of stop, I think this would be a good place.

\end{document}